\documentclass[a4paper,superscriptaddress,aps,prd,nofootinbib,floatfix,preprintnumbers,11pt]{revtex4-1}

\usepackage{enumerate}
\usepackage{amsmath,amssymb}
\usepackage{graphicx}
\usepackage{slashed}
\usepackage{xspace,slashed}
\usepackage{hyperref}
\hypersetup{colorlinks=true, citecolor=blue, urlcolor=blue, linkcolor=blue}
\usepackage[normalem]{ulem}
\usepackage{subfigure,orcidlink}
\usepackage{multirow,array}
\usepackage{siunitx}
\usepackage{listings}

\usepackage{hyperref}
\hypersetup{colorlinks=true, citecolor=blue, urlcolor=blue, linkcolor=blue}

\newcommand{\be}{\begin{eqnarray*}}
\newcommand{\ee}{\end{eqnarray*}}

\newcommand{\bee}{\begin{eqnarray}}
\newcommand{\eee}{\end{eqnarray}}
\newcommand{\beeq}{\begin{equation}}
\newcommand{\eeeq}{\end{equation}}

\newcommand{\beq}{\begin{eqnarray}} 
\newcommand{\eeq}{\end{eqnarray}}

\usepackage{titlesec}

\titleformat{\section}
  {\normalfont\Large\bfseries}{\thesection}{1em}{}
\titlespacing*{\section}
 {0pt}{1.ex plus .1ex minus .1ex}{2.ex plus .1ex minus .1ex}

\titleformat{\subsection}
  {\normalfont\bfseries}{\thesection.\thesubsection}{1em}{}
\titlespacing*{\subsection}
  {0pt}{1.ex plus .1ex minus .1ex}{0.ex plus .1ex minus .1ex}

\makeatletter
\renewcommand{\p@subsection}{\thesection.}
\renewcommand{\p@subsubsection}{\thesection.\thesubsection.}
\makeatother

\titleformat{\subsubsection}
  {\normalfont\itshape}{\thesection.\thesubsection.\thesubsubsection}{1em}{}
\titlespacing*{\subsubsection}
  {0pt}{1.ex plus .1ex minus .1ex}{0.ex plus .1ex minus .1ex}

\begin{document}
\allowdisplaybreaks
\flushbottom
\title{Distorting the Top Resonance with Effective Interactions}

\begin{abstract}
Interference effects in effective field theory (EFT) analyses can significantly distort sensitivity expectations, leaving subtle yet distinct signatures in the reconstruction of final states crucial for limit setting around Standard Model predictions. Using the specific example of four-fermion operators in top-quark pair production at the LHC, we provide a detailed quantitative assessment of these resonance distortions. We explore how continuum four-fermion interactions affect the resonance shapes, creating potential tensions between the high-statistics resonance regions and rare, high momentum-transfer continuum excesses. Our findings indicate that although four-fermion interactions do modify the on-shell region comparably to continuum enhancements, current experimental strategies at the High-Luminosity LHC are unlikely to capture these subtle interference-induced distortions. Nonetheless, such effects could become critical for precision analyses at future lepton colliders, such as the FCC-ee. Our work underscores the importance of resonance-shape measurements as complementary probes in global EFT approaches, guiding robust and self-consistent experimental strategies in ongoing and future high-energy physics programmes.
\end{abstract}
\author{Felix Egle\orcidlink{0009-0008-8205-2440}}\email{felix.egle@desy.de}
\affiliation{Deutsches Elektronen-Synchrotron DESY, Notkestr. 85, 22607 Hamburg, Germany\\[0.1cm]}
\author{Christoph Englert\orcidlink{0000-0003-2201-0667}}\email{christoph.englert@glasgow.ac.uk}
\affiliation{School of Physics and Astronomy, University of Glasgow, Glasgow G12 8QQ, United Kingdom\\[0.1cm]}
\author{Margarete M\"uhlleitner\orcidlink{0000-0003-3922-0281}}\email{margarete.muehlleitner@kit.edu}
\affiliation{Institute for Theoretical Physics, Karlsruhe Institute of Technology, 76128 Karlsruhe, Germany\\[0.1cm]}
\author{Michael Spannowsky\orcidlink{0000-0002-8362-0576}}\email{michael.spannowsky@durham.ac.uk}
\affiliation{Institute for Particle Physics Phenomenology, Department of Physics, Durham University, Durham DH1 3LE, United Kingdom\\[0.1cm]}

\allowdisplaybreaks
\setlength{\baselineskip}{0.9\baselineskip}

\maketitle
\preprint{DESY-25-044}
\preprint{IPPP/25/16}
\preprint{KA-TP-07-2025}

\section{Introduction}
\label{sec:intro}
Effective field theory (EFT)~\cite{Weinberg:1978kz,Georgi:1991ch} is increasingly becoming the new standard for framing the sensitivity to new physics interactions under well-defined theoretical assumptions. From the first proof-of-principle analyses, this has evolved into a cohesive programme spanning many different processes to obtain a global picture, including efforts by the experiments directly. Many approaches to setting constraints on process-relevant interactions rely on a good knowledge of relevant SM correlations to extract the SM null hypothesis used for setting~limits. 

When considering EFT deformations of the SM, this issue is further highlighted by the presence of non-resonant, non-SM contributions, which can lead to interference-related distortions of SM particle thresholds, e.g. through changing the line profile of intermediate unstable particles such as the top quark or the $W$ boson when reconstructed from their decay products. This can implicitly affect any EFT analysis at the Large Hadron Collider (LHC) and future facilities and needs to be contrasted with direct BSM-related modifications of the particles' resonance shapes directly.

\begin{figure}[!t]
\begin{center}
\includegraphics[width=0.42\textwidth]{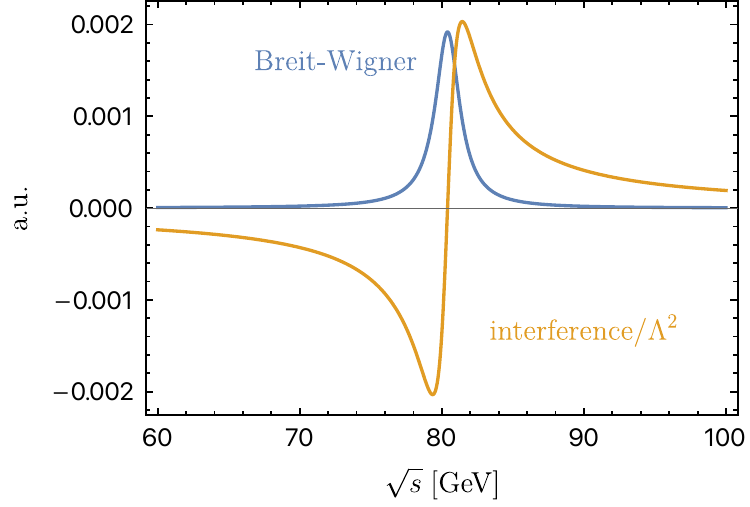}
\caption{\label{fig:example} Representative normalised Breit-Wigner distribution for the $W$ mass of $(m_W,\Gamma_W)\approx (80.4, 2.08)~\text{GeV}$. The Breit-Wigner suppression of the differential cross section is comparable with a loop suppression of a new physics contribution, which is assumed to be constant here.}
\end{center}
\end{figure}

Turning to analyses of top final states~\cite{Buckley:2015lku,Hartland:2019bjb,Brivio:2019ius,Bissmann:2020mfi,Ethier:2021bye,Garosi:2023yxg}, the relevance (and limitations) of top-pole measurements are well-documented in the literature. In the SM, non-perturbative effects~\cite{Beneke:1994sw,Bigi:1994em} are known to create systematic complications in extracting the top quark mass, chiefly from the analysis of top pair final states with a large abundance at the LHC. When turning to EFT modifications and, importantly, to new irreducible contributions to the amplitude of these final states, interference effects can enter the $pp\to \bar t t$ alongside the top quark decay. More concretely, the extraction of the top threshold that fundamentally underpins any signal-vs-background analysis of a $t\bar t$ system may be impacted by the presence of new physics contributions, e.g. by how much contact interactions in the vicinity of the $W$ threshold sculpt the kinematic correlations used to ``tag'' the top quark. Concretely, the squared SM-like matrix element in the vicinity of a resonance is well-described for widths $\Gamma \ll m$ by a Breit-Wigner distribution
\begin{equation}
{\cal{M}}_{\text{SM}} \underset{s\simeq m^2}{\simeq} {1\over s - m^2 + i\Gamma m}
\end{equation}
in the vicinity of the resonance mass for a relevant kinematic distribution $s$. Any additional, approximately constant contribution ${\cal{M}}_{\text{BSM}}$ for $s\simeq m^2$ therefore `tilts' the Breit-Wigner at interference-level $\text{d}\sigma \sim\, 2 \,\text{Re} ({\cal{M}}_{\text{SM}} {\cal{M}}_{\text{BSM}}^\ast)$. In doing so, the Breit-Wigner suppression can be numerically mitigated, however, it will decouple as a function of the SM width and the mass scale of the BSM theory $\Lambda$ , see Fig.~\ref{fig:example}.\footnote{Such effects have been discussed in the literature in the SM (for a recent review see e.g.~\cite{Dreiner:2008tw,Denner:2019vbn} including discussions of scheme dependencies).} The fact that such interference effects can severely impact and limit the experimental sensitivity to new physics is well-established in searches for exotic Higgs bosons~\cite{Gaemers:1984sj,Jung:2015gta,Hespel:2016qaf,Basler:2019nas}, which drives experimental searches, cf.~e.g.~\cite{ATLAS:2017snw}.

A phenomenologically relevant question therefore arises: Can effective contributions affect the reconstruction of SM candles which are used to set constraints on these interactions in the first place? If the answer to this question were yes, such modifications would enable a {\emph{process-specific}} test for new physics from reconstructed SM particle thresholds at the price of increased experimental complexity in defining the SM null hypothesis. A sizeable effect could also, in part, address the anomalous measurement of the $W$ mass as observed by the CDF collaboration~\cite{CDF:2022hxs} and open up new possibilities for new physics searches at future precision facilities, such as an FCC-ee.

Of course, any such new contribution might also be visible in the tails as a function of $s$. In the case of, e.g., four-fermion interactions, these effects are even kinematically enhanced. Therefore, it is not clear whether the limit setting from distribution tails is already constraining enough to render on-shell modifications (beyond total width and, hence, branching ratio fits) relevant. This note aims to reach a quantitative estimate of these effects within the SMEFT~\cite{Grzadkowski:2010es} framework. We focus on the $t\bar t$ final state as it is one of the most abundant processes at the LHC and, therefore, prone to provide good a priori sensitivity to SMEFT relevant deformations. We organise this work as follows: In Sec.~\ref{sec:simul}, we provide a short overview of our analysis setup before we consider representative effects in detail in Sec.~\ref{sec:analysis}. Section~\ref{sec:conc} provides a summary and conclusions with a positive message for the experimental community: We gather evidence that within the limits observed from tails, no significant on-shell distortion is observed. Current approaches can therefore be considered self-consistent and robust.

\section{Elements of the Analysis}
\label{sec:simul}
In this study, we use the SMEFT framework at dimension six~\cite{Grzadkowski:2010es}
\begin{equation}
{\cal{L}}={\cal{L}}_{\text{SM}} + \sum_i {c_i\over \Lambda^2} O_i \,,
\end{equation}
where $O_i$ denote the dimension-six operators and $\Lambda$ the new physics scale, to investigate the influence of new physics on top quark pair production at the LHC. In our analysis, we isolate the contributions from new EFT physics by adjusting the Wilson coefficients $c_i$. Only one Wilson coefficient was set to a non-zero value at a time (which enables a full fit at interference level between $\langle{\cal{L}}_{\text{SM}}\rangle$ and $\langle O_i\rangle$), allowing us to disentangle the effects of individual operators. For the generation of our event samples we employ {\tt{MadGraph\_aMC@NLO}}~\cite{Alwall:2014hca} together with {\tt{SMEFTsim3.0.2}}~\cite{Brivio:2020onw}. In practice, we generate around 2 million events for the SM as well as 1 million events for the EFT operator interference (we comment on the so-called quadratic dimension-six effects below), which is enough statistics for a smooth extrapolation to the high-luminosity phase of the LHC (3/ab) whilst capturing subtle deviations with sufficient statistics.

To highlight the on-shell/off-shell interplay alluded to above, we focus firstly on the operators
\begin{align}
Q^{(3)}_{lq} &= (\bar l \gamma_\mu \tau^I l_R) \; (\bar q_3 \gamma^\mu\tau^I q_3),
\end{align}
which can interfere with the top decay mediated by the $W$ boson in the SM. As the $W$ boson decays on-shell, the $Q^{(3)}$-interference term distorts the shape of the $W$ resonance that, in turn, is used in generic reconstruction techniques for the top quark~\cite{ATLAS:2022xfj}. 

In this note, we do not attempt to establish a global picture of top-quark interactions but note that the new physics operators used here as a `strawman' scenario are also relevant in flavour physics, see, e.g.~\cite{Chala:2018agk,Bissmann:2019gfc,Grunwald:2023nli,Bissmann:2020mfi}. 

\begin{figure*}[!t]
\begin{center}
\parbox{0.49\textwidth}{
\includegraphics[width=0.49\textwidth]{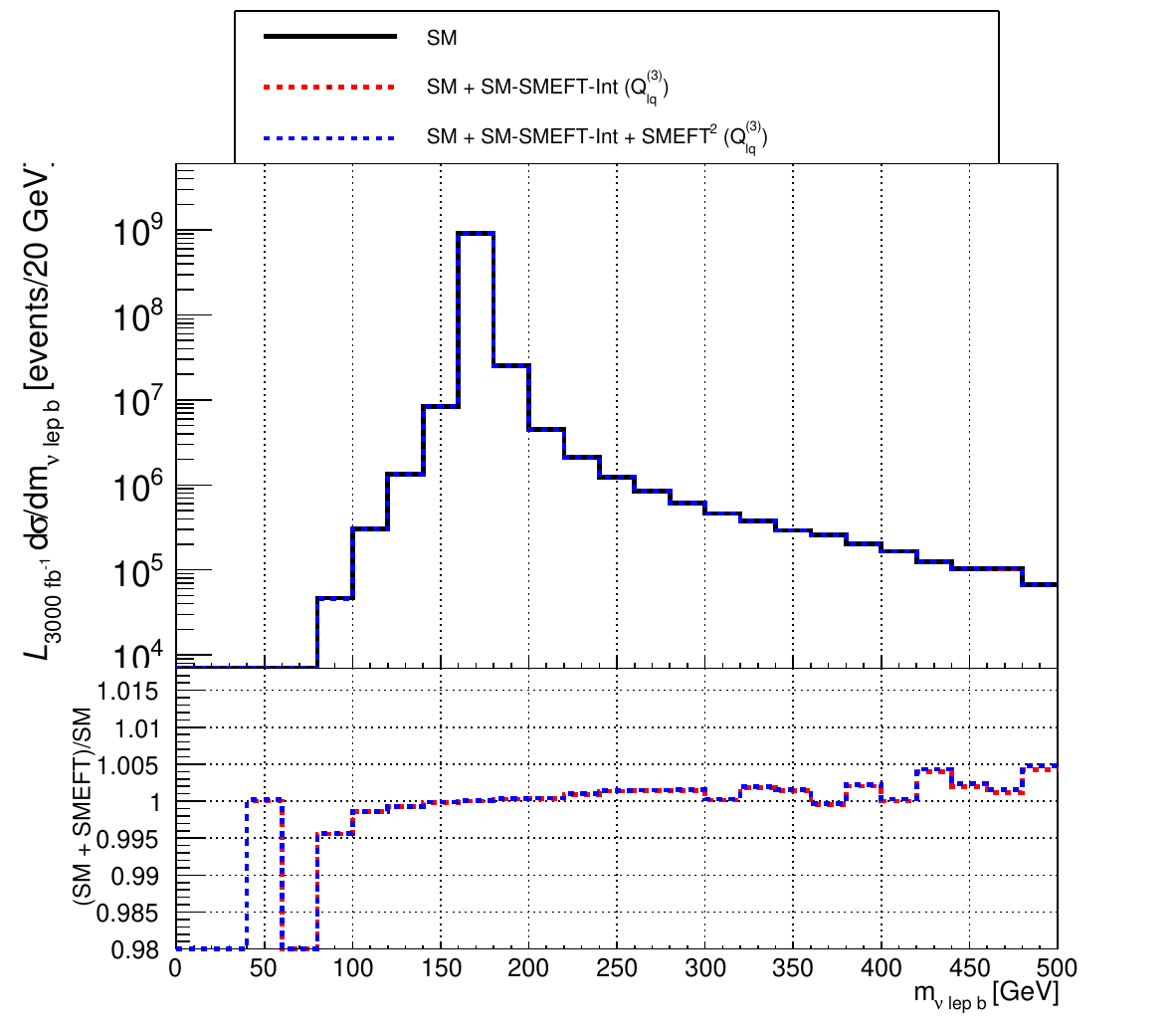}
}
\hfill
\parbox{0.49\textwidth}{
\includegraphics[width=0.49\textwidth]{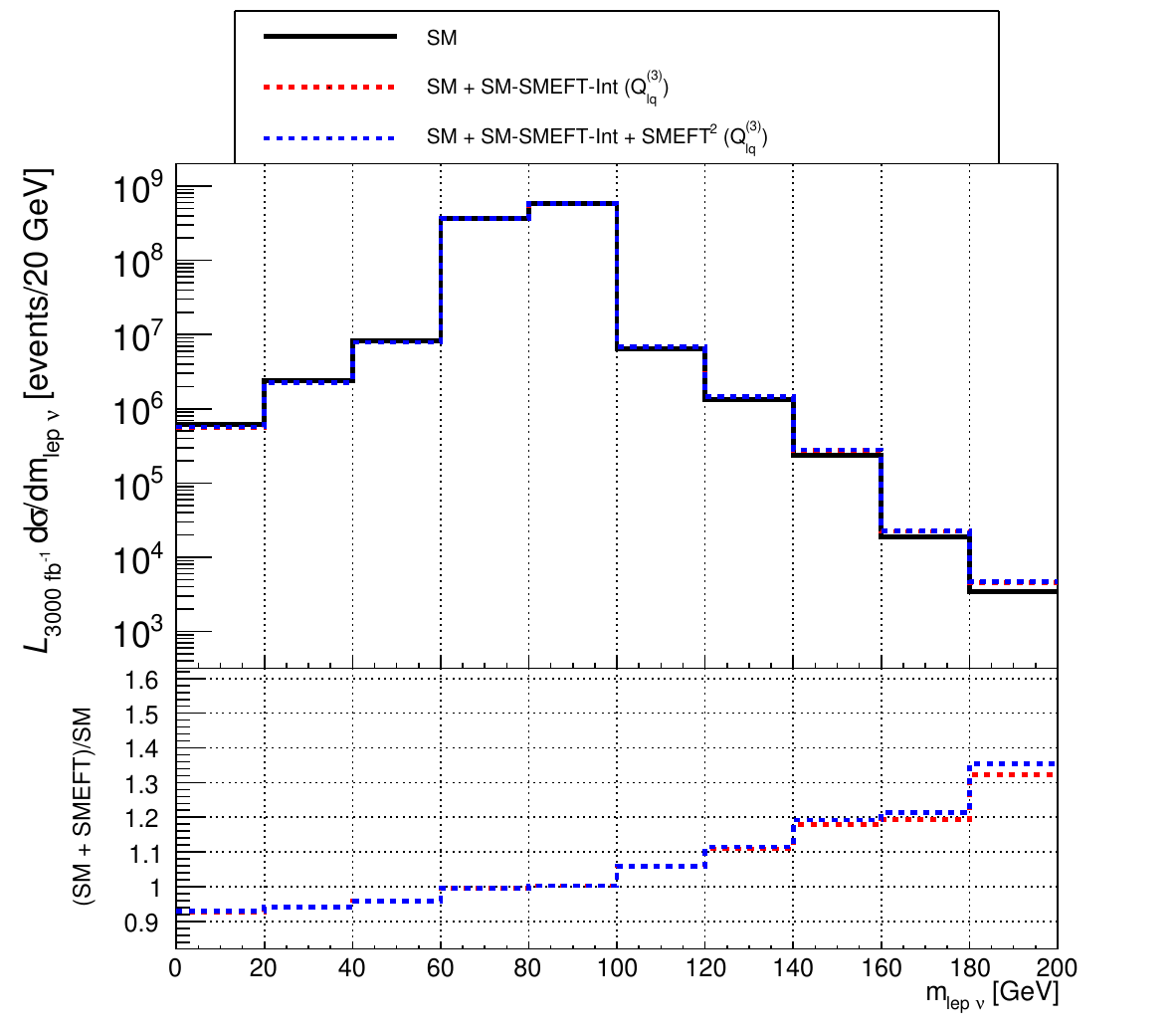}
}
\parbox{0.49\textwidth}{
\includegraphics[width=0.49\textwidth]{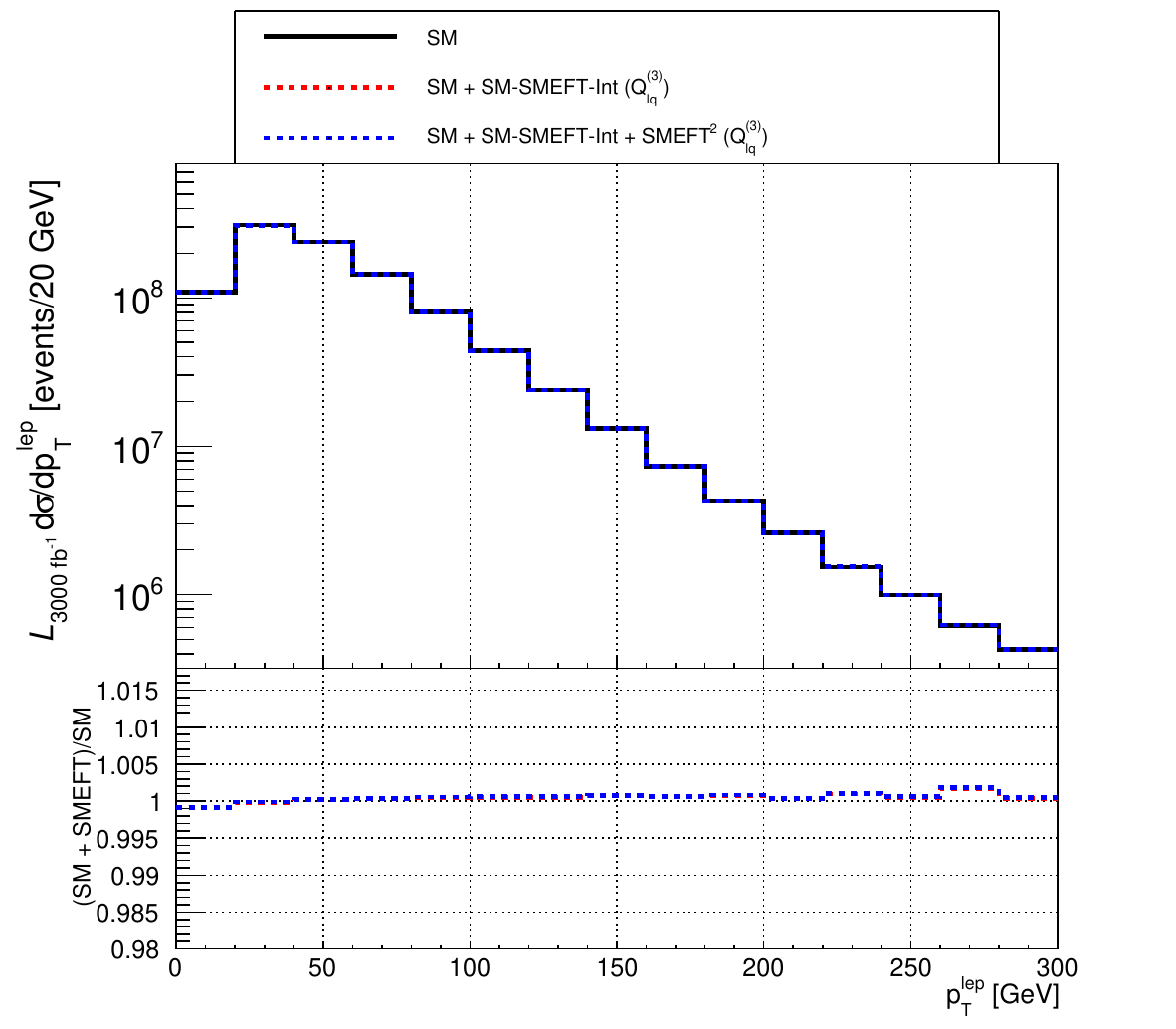} 
}
\hfill
\parbox{0.49\textwidth}{\vspace{0.5cm}
\caption{
\label{fig:clq3} Energy-transfer sensitive differential distributions of semi-leptonic $t\bar t$ LHC production with a priori sensitivity to the new interactions discussed in this work, here specifically for the operator $O^{(3)}_{lq}$ (with strength $1/\text{TeV}^2$) for illustration purposes. Particularly relevant to the focus of this work are the invariant mass distributions, which are sensitive to resonance distortion. The lower inserts show the ratio of BSM distributions (either at dimension-six level or including squared dimension-six contributions) with respect to the SM.
}}
\end{center}
\end{figure*}
We analyse a range of differential distributions, such as the transverse momentum (\( p_T \)) for each final state particle and various invariant mass systems linked to the top-quark decay. We conservatively take bin sizes of 20 GeV and focus on fully and semi-leptonic top decays from $pp\to t\bar t$ collisions at 13~TeV. The selection of different bin sizes allows us to study the results' sensitivity to the data's granularity when considering the impact of the aforementioned interference effects. To quantify the deviations from the Standard Model predictions, we performed a $\chi^2$ test, defined as
\begin{equation}
\label{eq:chi2test}
\chi^2 = \sum_{i}^{\text{\# bins}} \frac{(n_{i,\text{SM+SMEFT}} - n_{i,\text{SM}})^2}{\sigma_{i,\text{SM+SMEFT}}^2 + \sigma_{i,\text{SM}}^2}
\end{equation}
where \( \sigma_i \) is the error on a bin count $n_i=\sigma_i\times {\cal{L}}$, given by \( \sqrt{n_i} \) (we comment on the impact of systematic uncertainties below). This method allows us to assess the goodness of fit between the SM predictions and the observed data, including potential contributions from new physics. To evaluate the critical threshold above which the SM null hypothesis will be rejected (and to compute 95\% confidence level limits depending on the degrees of freedom relevant to the specific distribution under investigation), we employ a bisection method to find the limits for the Wilson coefficients, which iteratively adjusts the coefficients to find the point where the observed \( \chi^2 \) exceeds the critical value (determined by the number of bins and confidence level).

Of course, four-fermion interactions involving a single top quark are already constrained from top width measurements, i.e. fits to the top quark's branching ratios, but also flavour physics measurements~\cite{Bissmann:2019gfc,Bissmann:2020mfi,Grunwald:2023nli}. To reflect the measurements directly relevant to our analysis~\cite{Chala:2018agk,Atkinson:2024hqp}, we include the top width measurement of $\Gamma_t =1.34\pm 0.16~\text{GeV}$~\cite{ParticleDataGroup:2024cfk} as an additional contribution to the $\chi^2$ test statistic. To gain a quantitative understanding of how operators can lead to on-shell distortion in tension against constraints from continuum enhancement, we compare the limits of two different exclusive phase space regions
\begin{enumerate}[a)]
\item the ``on-shell region'' related to the top and $W$ thresholds, respectively. We include all bins around the $t$ and $W$ threshold in the corresponding invariant mass distributions that are in the region $[m_t-2\Gamma_t,m_t+2\Gamma_t]$ and $[m_W-2\Gamma_W,m_W+2\Gamma_W]$.
\item the ``off-shell region'' that complements the on-shell regions as defined above; in particular, these contain the tails of the distributions susceptible to EFT continuum modifications. More precisely, this selects the tail region, only considering bins beyond the OS region.
\end{enumerate}
The interference-related histograms for $O_{lq}^{(3)}$ are shown in Fig.~\ref{fig:clq3}, for momentum-dependent distributions that are used both for the reconstruction of the signal process and the new physics limit setting, see e.g.~\cite{ATLAS:2022xfj}.

\section{Relevance of Resonance Distortion from a Toy Fit}
\label{sec:analysis}
Any significant distortion of the involved resonance structures already at the Monte Carlo truth indicates a significant a priori issue for experimental analyses. From Fig.~\ref{fig:clq3} it is immediately obvious that not all observables are equally suited to set strong constraints on the presence of new physics even when they are correlated. For instance, for contact interaction that is related to the leptonic top decay, there are cancellations in the reconstructed top mass that are most efficiently resolved in the $W$ mass. The behaviour of the $l\nu$ system below and above the top threshold is exactly the behaviour motivated in Fig.~\ref{fig:example}, coarse-grained to the bin size of 20~\text{GeV} that we consider in this work (this seems a conservative choice in the light of Ref.~\cite{ATLAS:2023gsl} and the data improvement that we can expect at the high-luminosity LHC phase).

\begin{figure}[!t]
\centering
\includegraphics[width=0.48\textwidth]{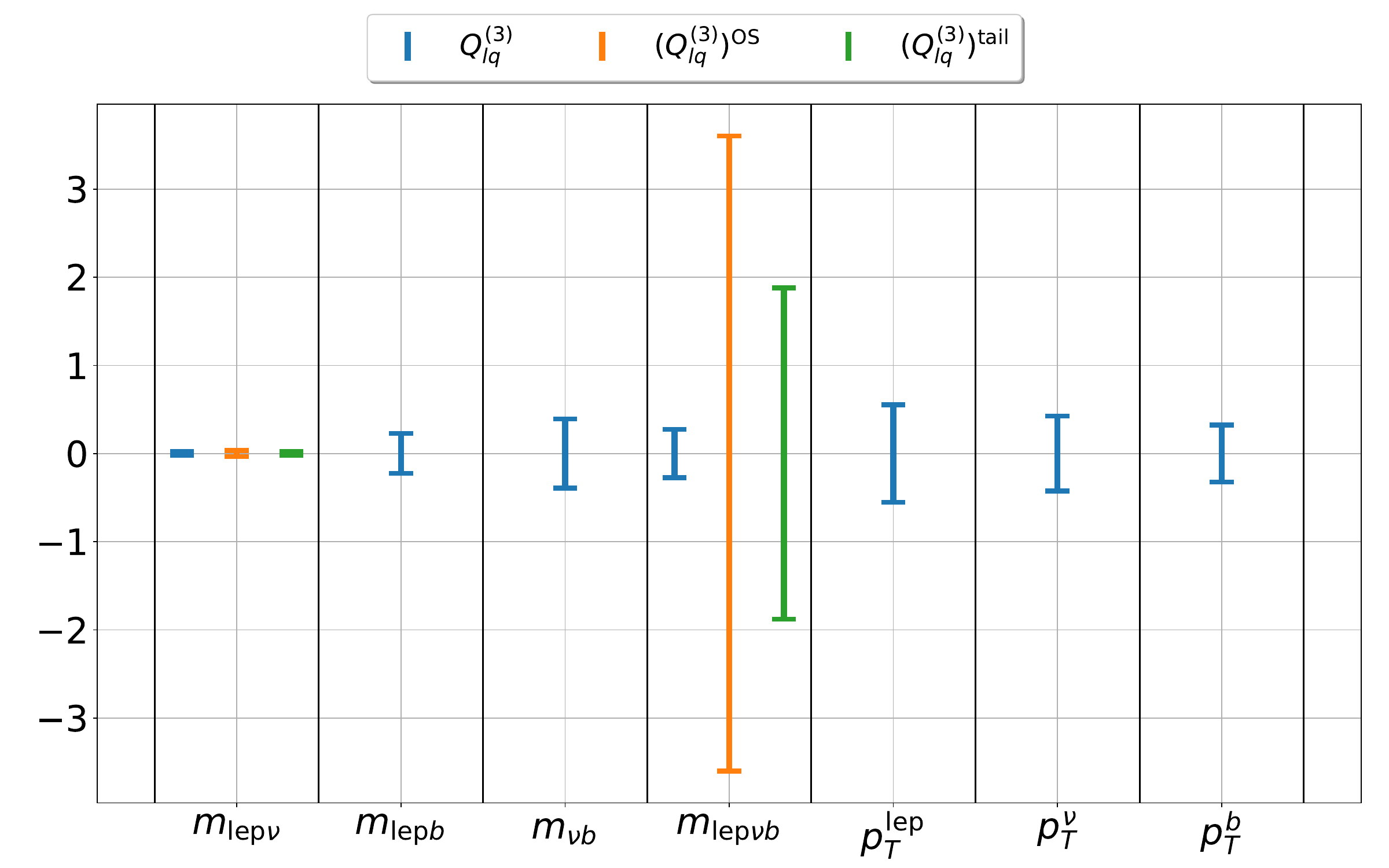} 
\caption{HL-LHC constraints on the four-fermion interactions discussed in this work for $\Lambda=1~\text{TeV}$, including a comparison of the single discriminants for the limit setting. Where we consider on-shell and tail distributions, the different constraints are highlighted in orange and green, respectively. No systematics are included in this comparison; for additional details, see the main body. \label{fig:clq3qq1lim}}
\end{figure}

The result of the described limit setting exercise is shown in Fig.~\ref{fig:clq3qq1lim}. Of course, systematics can further impact the quality of the limit extraction. If they are included as a fractional contribution related to the bin entries in Eq.~\eqref{eq:chi2test}, e.g., the limit on $Q_{lq}^{(3)}$ from the $p_\mathrm{T}^b$ distribution changes from $[-0.323, 0.323]$ to $[-0.404,0.404]$ for a (conservative) systematic uncertainty of $25\%$ ($\Lambda=1~\text{TeV}$). Similarly, the limit on $Q_{lq}^{(3)}$ using the $m_{\mathrm{lep}\nu}$ distribution changes from $[-0.021, 0.021]$ to $[-0.026, 0.026]$.

With these results at hand, we can conclude that resonance distortion is not a relevant effect in the hadron collider environment, and the SM reconstruction techniques remain robust: Both the OS and the non-resonant region might provide comparable statistical sensitivity with interference effects present, but the coarse-graining of measured invariant mass distributions mitigates distortion effects leaving only a slight asymmetry as visible in Fig.~\ref{fig:clq3}. This still could leave them relevant for unbinned approaches; we encourage the LHC experiments to include self-consistency checks along the lines of the discussion in this note. 

\begin{figure*}[!b]
\begin{center}
\includegraphics[width=0.49\textwidth]{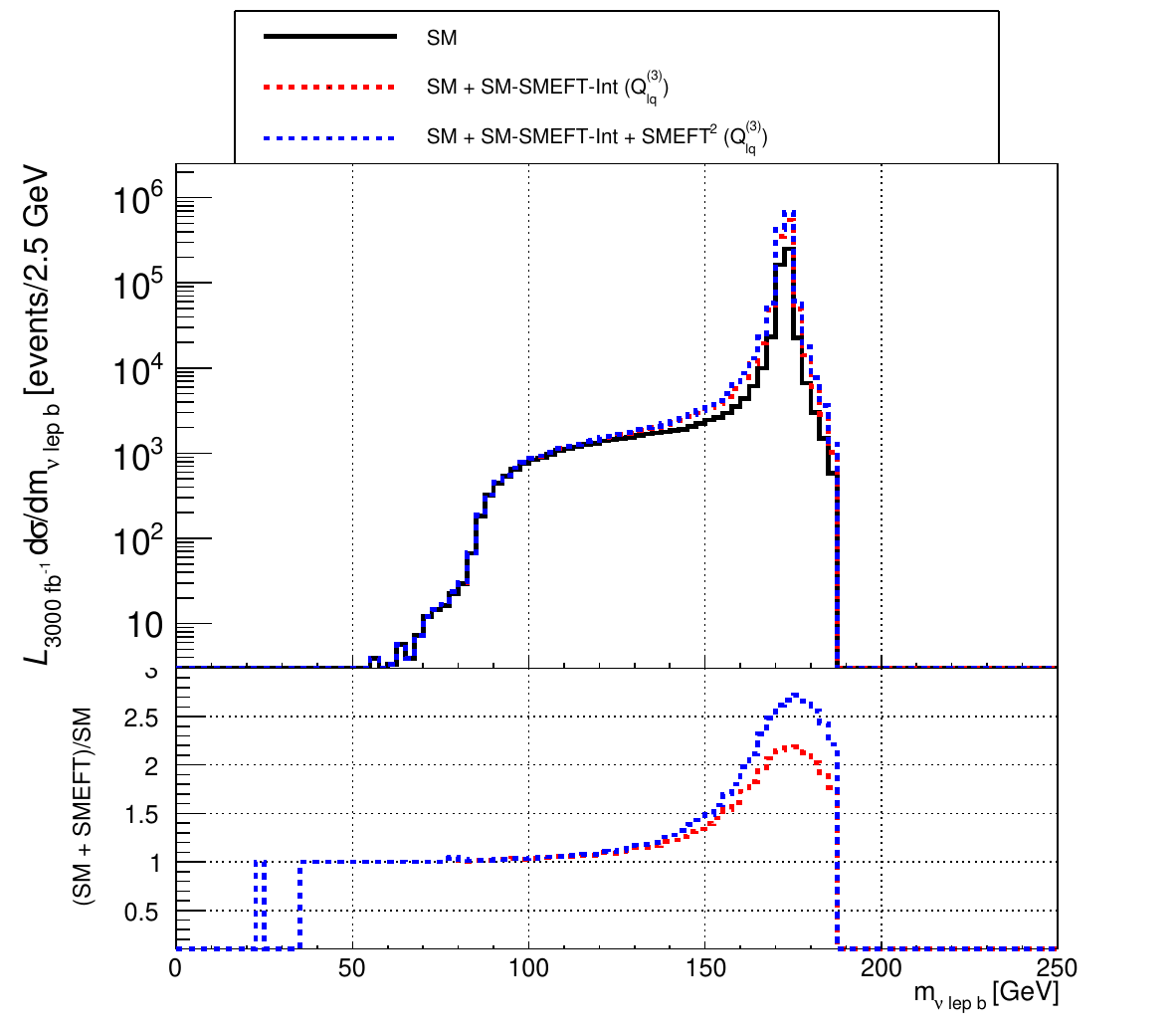} 
\hfill
\includegraphics[width=0.49\textwidth]{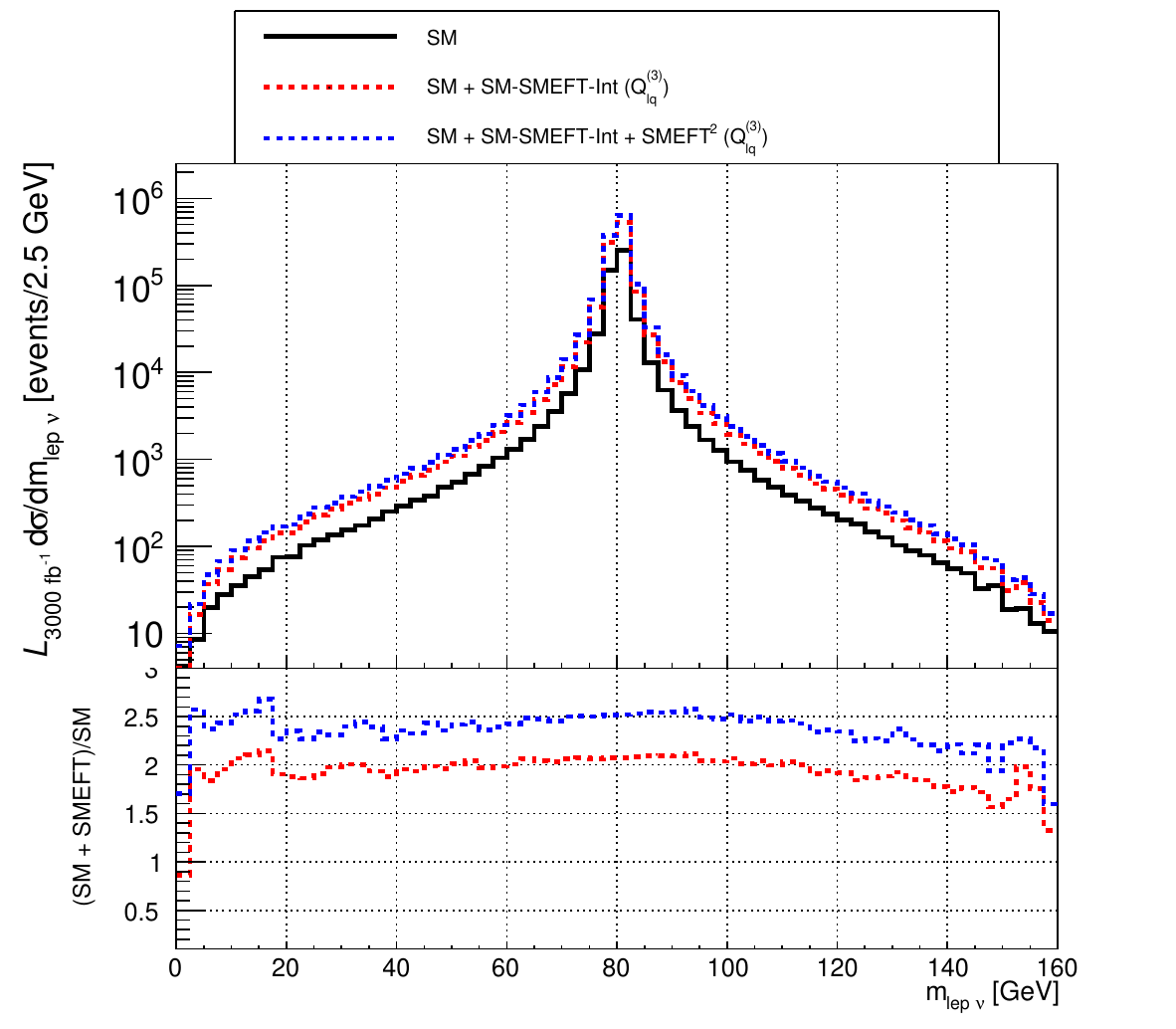}  
\caption{\label{fig:clq3mmevb_e+e-} $W$ lineshape-related observables (again with strength $1/\text{TeV}^2$ for illustration) with a priori sensitivity to resonance distortion for $e^+ e^-$ collisions running close to the $t\bar t$ threshold. The shape change of the distributions relative to the SM is more pronounced than at hadron colliders, indicating a greater intrinsic sensitivity at the FCC-ee.
}
\end{center}
\end{figure*}

In collider environments that enable a more detailed exploration of the top pair threshold, this could, in principle, be different. A potential FCC-ee, which is under active consideration as part of the current ECFA process, could feature a top quark pair production programme at a c.m.~energy of $\sqrt{s}=350-365$~GeV, collecting a luminosity of around $1.8 \times 10^4~{\text{fb}}^{-1}$ per year. To gain a quantitative estimate of the relevance of the effects discussed in the work in such an environment, we re-perform the above limit setting analysis for $e^+ e^- \to t\bar t$, running at the top pair threshold (see also~\cite{Fadin:1987wz,Hoang:1998xf,Hoang:1999zc,Hoang:2001mm,Blondel:2021ema} and the BSM studies of \cite{Khiem:2015ofa,Janot:2015yza,Englert:2017dev,Jafari:2019seq,Ma:2024izj}). More concretely, we consider a centre-of-mass energy of $2 \times {180}~\text{GeV}$ and a luminosity of ${\cal{L}}=2.56~\text{ab}^{-1}$~\cite{CLICdp:2018esa}\footnote{See also the resources of the \href{https://hep-fcc.github.io/FCCeePhysicsPerformance/}{FCC-ee Physics Performance Group}.}.

\begin{figure}[!t]
\centering
\includegraphics[width=0.49\textwidth]{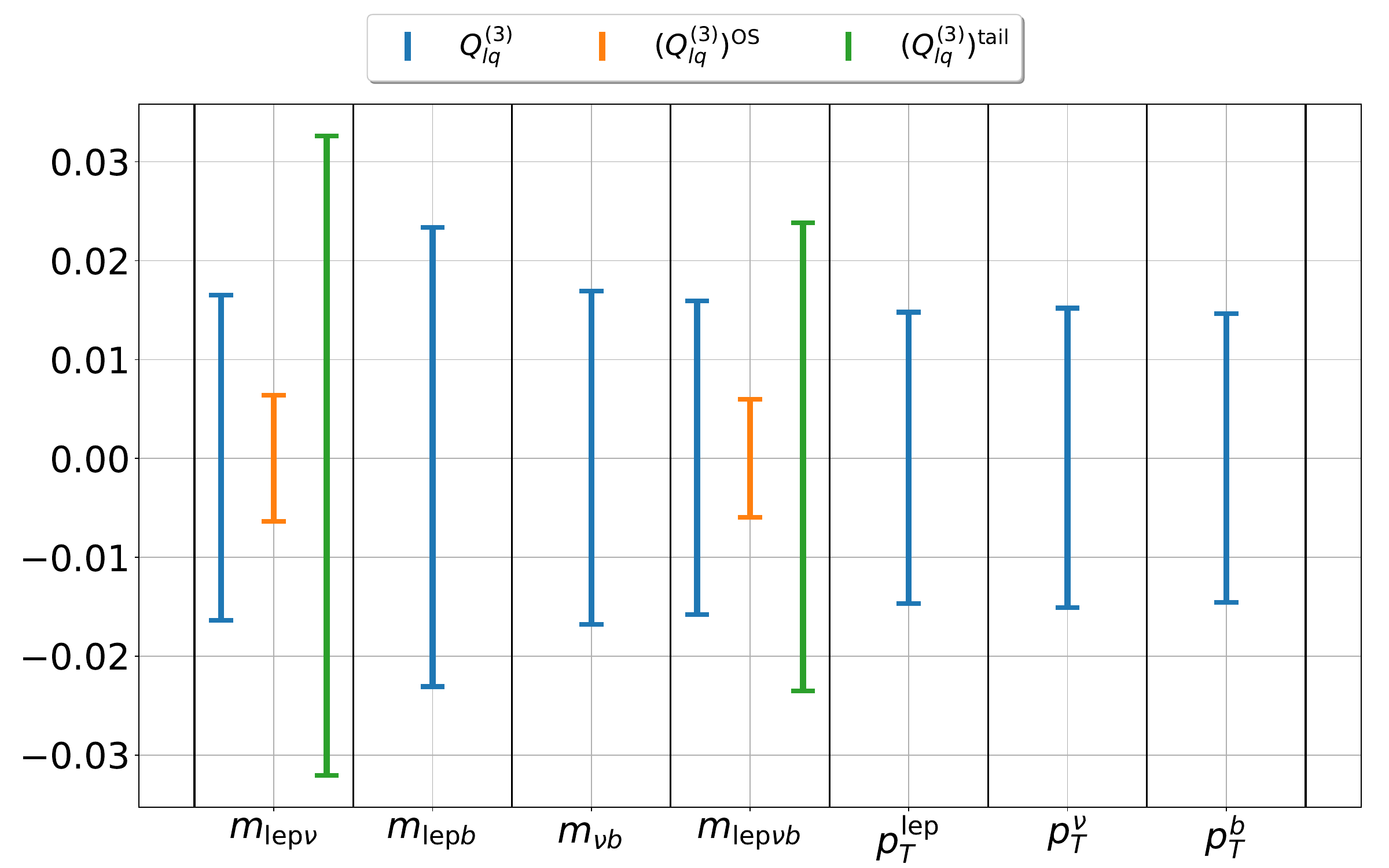} 
\caption{Limits on $O_{lq}^{(3)}$ (in units $1/\text{TeV}^2$) at an $e^+e^-$ collider running close to the $t\bar t$ threshold with an integrated luminosity of ${\cal{L}}=2.56~\text{ab}^{-1}$. The impact of systematics is neglected but commented on in the text. \label{fig:clq3lime+e-}}
\end{figure}

Comparing results to the HL-LHC environment, we again observe a slight asymmetry of the reconstructed $W$ mass in $t\bar t$ final states, see Figs.~\ref{fig:clq3mmevb_e+e-},~\ref{fig:clq3lime+e-}. The precision environment of a lepton collider enables more fine-grained measurements, and we opt for a 2.5~GeV binning. In particular, we see that observables that are used in experimental threshold analyses show a quantitatively modified behaviour. The limit from the $m_{\mathrm{lep}\nu}$ distribution considering only the OS region for $Q_{lq}^{(3)}$ is  $[-0.0059, 0.0059]$ (decreasing to $[-0.0074, 0.0074]$ for a flat 25\% systematic uncertainty); from the $p_\mathrm{T}^b$ distribution we obtain $[-0.015, 0.015]$ $([-0.018, 0.018])$. However, it is worth pointing out that these effects are also susceptible to the theoretical uncertainties related to renormalisation scheme choices. We can also conclude that, for this environment, a continuum BSM contribution will unlikely assert itself dominantly through a production mechanism-dependent reconstruction difference of the $W$ boson in $t\bar t$ production, cf.~Fig.~\ref{fig:clq3mmevb_e+e-}. Of course, a wider range of interactions is relevant for an agnostic search for non-resonant physics.

\section{Conclusions}
\label{sec:conc}
The study presented here highlights and quantifies the effects of EFT modifications on resonance shapes in top-quark pair production, with a particular emphasis on four-fermion operators within the SMEFT framework. Our detailed investigation of key kinematic distributions relevant to the reconstruction of top-quark events at the high-luminosity LHC reveals subtle interference effects that distort resonance shapes. However, these interference-induced resonance distortions remain negligible due to current collider experimental resolutions and statistical limitations. Thus, existing analysis methodologies at the HL-LHC are robust and are not compromised by the subtle effects explored in this study.

Nevertheless, the potential impact of these resonance distortions should not be dismissed, particularly in the context of future collider environments offering significantly enhanced sensitivity. For example, resonance distortions can become observable at an electron-positron collider like the FCC-ee, operating near the top-quark production threshold with higher resolution precision elect. Our projections indicate tighter constraints on EFT parameters, i.e. we find a refined sensitivity range of [-0.0059, 0.0059] obtained from analysing the invariant mass distribution of leptonic $W$ decays at the FCC-ee. This level of sensitivity surpasses what is achievable at the current LHC and shows the value of future precision experiments (see also~\cite{Celada:2024mcf}).

\bigskip 
\noindent {\bf{Acknowledgements}} ---
A Leverhulme Trust Research Project Grant RPG-2021-031 funds this work. 
F.E. is supported by the DFG Emmy Noether Grant No.~BR 6995/1-1. F.E.~acknowledges support by the Deutsche Forschungsgemeinschaft (DFG, German Research Foundation) under Germany's Excellence Strategy --- EXC 2121 ``Quantum Universe'' --- 390833306. This work has been partially funded by the Deutsche Forschungsgemeinschaft (DFG, German Research Foundation) --- 491245950.
C.E. is supported by the UK Science and Technology Facilities Council (STFC) under grant ST/X000605/1 and the Institute for Particle Physics Phenomenology Associateship Scheme.
M.M. is partly supported by the BMBF-Project 05H21VKCCA and acknowledges partial support by the Deutsche Forschungsgemeinschaft (DFG, German Research Foundation) under grant 396021762 - TRR 257.  
M.S. is supported by the STFC under grant ST/P001246/1.

\bibliography{references}

\end{document}